\documentclass[a4paper,11pt]{article}
\usepackage{latexsym}
\usepackage{chapterbib}
\usepackage{graphics,graphicx}
\usepackage{amssymb,amsfonts,amstext}
\usepackage[tbtags]{amsmath}
\usepackage{hyperref}
\usepackage{color,xcolor}
\usepackage{hyperref}
\usepackage{geometry}  
\usepackage{braket}
\usepackage{cancel}
\usepackage[utf8]{inputenc}
\usepackage{datetime}
\usepackage{slashed}
\usepackage{tikz}
\usepackage{dsfont}
\usepackage{amsthm}
\usepackage{bm}
\usepackage{tabu}
\usepackage{cite}

\newcommand{\llangle}{\langle \langle}
\newcommand{\rrangle}{\rangle \rangle}

\allowdisplaybreaks

\newcommand\hcancel[2][black]{\setbox=\hbox{$#2$}%
	\rlap{\raisebox{.45\ht0}{\textcolor{#1}{\rule{\wd0}{1pt}}}}#2} 

\def\){\right)}
\def\({\left( }
\def\]{\right] }
\def\[{\left[ }

\def\NO{\nonumber}

\newcommand{\be}{\begin{equation}}
\newcommand{\ee}{\end{equation}}

\def\bea{\begin{eqnarray}}
\def\eea{\end{eqnarray}}

\def\bal#1\eal{\begin{align}#1\end{align}}
\def\bala#1\eala{\begin{align*}#1\end{align*}}

\newcommand\numberthis{\addtocounter{equation}{1}\tag{\theequation}}

\def\bald{\begin{aligned}}
\def\eald{\end{aligned}}

\def\bsub{\begin{subequations}}
\def\esub{\end{subequations}}
\def\bsal#1\esal{\begin{subequations}\begin{align}#1\end{align}\end{subequations}}

\def\beqx{\begin{displaymath}}
\def\eeqx{\end{displaymath}}

\newcommand{\bmat}{\left(\begin{array}}
\newcommand{\emat}{\end{array}\right)}

\def\a{\alpha}
\def\b{\beta}
\def\c{\chi}
\def\d{\delta}
\def\e{\epsilon}
\def\f{\phi}
\def\g{\gamma}
\def\h{\eta}
\def\j{\psi}
\def\k{\kappa}
\def\l{\lambda}
\def\m{\mu}
\def\n{\nu}
\def\o{\omega}

\def\r{\rho}
\def\s{\sigma}

\def\x{\xi}
\def\z{\zeta}
\def\D{\Delta}
\def\F{\Phi}
\def\G{\Gamma}

\def\L{\Lambda}
\def\O{\Omega}

\def\ve{\varepsilon}

\def\ca{{\cal A}}
\def\cb{{\cal B}}
\def\cc{{\cal C}}
\def\cd{{\cal D}}

\def\cf{{\cal F}}

\def\cj{{\cal J}}
\def\ck{{\cal K}}
\def\cl{{\cal L}}
\def\cm{{\cal M}}
\def\cn{{\cal N}}
\def\co{{\cal O}}

\def\cq{{\cal Q}}
\def\car{{\cal R}}
\def\cs{{\cal S}}
\def\ct{{\cal T}}

\def\bo{{\raise-.3ex\hbox{\large$\Box$}}}               
\def\pa{\partial}                                       
\def\face{{\raise.2ex\hbox{$\displaystyle \bigodot$}\mskip-2.2mu \llap {$\ddot
        \smile$}}}                                   
\def\>{\rangle}                                      
\def\<{\langle}                                      

\def\leftrightarrowfill{$\mathsurround=0pt \mathord\leftarrow \mkern-6mu
        \cleaders\hbox{$\mkern-2mu \mathord- \mkern-2mu$}\hfill
        \mkern-6mu \mathord\rightarrow$}        
\def\dvec#1{\vbox{\ialign{##\crcr
        \leftrightarrowfill\crcr\noalign{\kern-1pt\nointerlineskip}
        $\hfil\displaystyle{#1}\hfil$\crcr}}}           

\def\-{\hphantom{-}}
\newcommand{\dd}{\mbox{d}}

\title{Quantum consistency of rigid supersymmetric theories}

\begin{document}

\begin{titlepage}

\pagestyle{empty}

\begin{flushright}
 \end{flushright}
\vskip1.5in

\begin{center}

{\bf  \Large 
Quantum consistency in supersymmetric  theories with $R$-symmetry in curved space }
\end{center}
\vskip0.2in

\begin{center}
{ \large Ok Song An,$^{a}$ 
Jin U Kang,$^{b,a}$ Jong Chol Kim,$^{a}$ Yong Hae Ko$^{a}$}
\end{center}
\vskip0.2in

\begin{center}
{ {$^{a}$}\emph{ Department of Physics, Kim Il Sung University \\ Ryongnam Dong, TaeSong District, Pyongyang, DPR Korea}}\\ \vskip0.1in
{ {$^{b}$} \emph{International Centre for Theoretical Physics  \\ Strada Costiera 11, 34151 Trieste, Italy} 
}\\
\vskip0.1in
\end{center}
\vskip0.2in

\begin{abstract}

We discuss consistency  at the quantum level 
in the rigid  $\mathcal N=1$ supersymmetric field theories with a $U(1)_R$ symmetry in four-dimensional curved space which are formulated via coupling to the new-minimal supergravity background fields. 
By analyzing correlation functions of the current operators in the  $\mathcal{R}$-multiplet, we show that
the quantum consistency with the (unbroken) supersymmetry requires the $U(1)_R$ anomaly coefficient, which 
depends only on the field content of the theory,  to vanish. This consistency condition is obtained under the assumption that the supercurrent Ward identity is non-anomalous and that the vacuum is supersymmetric.  

\end{abstract}

\end{titlepage}

\tableofcontents
\addtocontents{toc}{\protect\setcounter{tocdepth}{3}}
\renewcommand{\theequation}{\arabic{section}.\arabic{equation}}

\section{Introduction}
\setcounter{equation}{0}

Supersymmetric field theories in curved space have attracted many interests in recent years. For such theories supersymmetric localization technique \cite{Nekrasov:2002qd,Pestun:2007rz} allows the non-perturbative exact computation of some interesting physical quantities such as the partition function and BPS Wilson loops, which 
can be used to test the duality conjectures like the AdS/CFT correspondence \cite{Maldacena:1997re,Witten:1998qj,Gubser:1998bc}. In this paper we focus on supersymmetric field theories with a $U(1)_R$ symmetry on 
curved manifolds in 3+1 dimensions.

According to \cite{Festuccia:2011ws,Dumitrescu:2012ha, Klare:2012gn} (see also \cite{Dumitrescu:2016ltq} for a recent review), one can formulate a $\cn=1$ theory with a $U(1)_R$ symmetry in 4D curved space via coupling to the  new-minimal supergravity \cite{Sohnius:1981tp,Sohnius:1982fw}: One first couples the $\car$-multiplet to the  new-minimal supergravity multiplet  and then take a \emph{rigid} limit sending the Newton's constant to zero, so that the supergravity is decoupled while the fields in the supergravity multiplet are sent to fixed backgrounds. In constructing supersymmetric field theories in curved space, background fields are typically chosen to be bosonic, and consistency with supersymmetry requires the supersymmetry variation of the gravitino in the gravity multiplet to vanish, which leads to a \emph{generalized Killing spinor} (GKS) equation. 
For each solution of the GKS equation there exists a conserved supercharge. In the case of 4D $\cn=1$ theory with a $U(1)_R$ symmetry, the GKS equation can have a solution if and only if the background manifold admits an integrable complex structure and a compatible Hermitian metric.

Important exact results were obtained for the 4D $\cn=1$ supersymmetric field theory with an $R$-symmetry in \cite{Closset:2013vra,Closset:2014uda,Assel:2014paa,Cassani:2013dba, Assel:2015nca,Cassani:2014zwa}, using localization technique. Here we list some of them:
\begin{itemize}
\item When there are two supercharges with opposite $U(1)_R$ charge, supersymmetric renormalization scheme is free of ambiguities and the partition function is invariant under the deformation of the Hermitian metric for a fixed complex structure.
\item When the background manifold is $S_{r_3}^3\times S_\b^1$, the supersymmetric Casimir energy becomes
\begin{equation}\label{eqn:e_susy}
E_{\rm susy}\equiv -\lim_{\b\rightarrow\infty}\frac{\dd}{\dd\b}\log Z_{\rm susy}=\frac{4}{27r_3}(a+3c),
\end{equation}
where $a$ and $c$ are two trace anomaly coefficients in four dimensions.
\end{itemize}

We recall that the field theory computations in \cite{Closset:2013vra,Closset:2014uda,Assel:2014paa,Cassani:2013dba, Assel:2015nca,Cassani:2014zwa} 
were carried out by using the supersymmetric Ward identities and the supersymmetry algebra  which are 
derived from the \emph{classical} new-minimal supergravity. 
In \cite{Gates:1981yc} (see also section 7.10 of \cite{Gates:1983nr}) it was argued via the superspace formalism that the new-minimal supergravity 
can be in general quantum-mechanically inconsistent, due to the appearance of the ``superscale'' anomalies (consisting of the conformal anomaly, 
the $U(1)_R$ chiral anomaly and the $\gamma$-trace of the supercurrent) that are  inconsistent with the local supersymmetry Ward identities. 
Then one could also question the quantum consistency of the new-minimal supergravity formulation of the rigid 
 $\cn=1$ field theories with an $R$-symmetry.     
However, \cite{Gates:1981yc} considered only the one-point function of the supercurrent superfield and 
one could expect that there is no inconsistency for the backgrounds on which the anomalies in the Ward identities (for current one-point functions) are \emph{numerically} vanishing. 
In particular, the backgrounds considered in \cite{Assel:2014paa, Cassani:2013dba ,Assel:2015nca,Cassani:2014zwa,Genolini:2016sxe,Genolini:2016ecx} are anomaly-free in this sense. 
Nevertheless, the anomalies might appear in the higher-point correlation functions as the \emph{contact terms}, some of which may be physically meaningful. 
This motivates us to study the higher-point correlation functions to investigate the quantum consistency.    

In this paper we analyze the two-point and higher-point correlation functions
of the current operators in the $\car$-multiplet.  
Assuming that the supercurrent Ward identity is non-anomalous and that the vacuum is supersymmetric, we show that the unbroken supersymmetry is  inconsistent at the quantum level unless the coefficient of the $U(1)_R$ anomaly vanishes.
Note that the anomaly coefficients depend only on the field content of the theory. 
Since the Ward identities and the rigid supersymmetry algebra are a direct consequence of the new-minimal supergravity, 
this implies that the 4D $\cn=1$ field theory with an $R$-symmetry can be consistently formulated in terms of the new-minimal supergravity only for some special systems 
with the field content that leads to vanishing coefficient of $U(1)_R$ anomaly. One example of this is  
a theory that consists of a free chiral multiplet with $R$-charge 1, since in this case the Weyl fermion in the chiral multiplet is uncharged under the $U(1)_R$ symmetry and gives no contribution to the $U(1)_R$ chiral anomaly.

The rest of this note is organized as follows. In section \ref{sec:nm-sugra} we briefly review the symmetries of the new-minimal supergravity, the definition of the generalized Killing spinor (GKS) and the construction of the Killing vector from the GKSs. We then derive the Ward identities of $\cn=1$ supersymmetric field theories with an $R$-symmetry in section \ref{sec:ward}, the results of which are used to reproduce the rigid supersymmetry algebra in section \ref{SUSY}. In section \ref{sec:quantum-consistency} we show that in order for $\cn=1$ field theories with an $R$-symmetry to be consistently formulated via the new-minimal supergravity the (pure) $U(1)_R$ chiral anomaly coefficient should vanish. 
Finally, we end with concluding remarks in section \ref{sec:discussion}.  
Appendix \ref{app:conventions} contains our conventions, while appendix \ref{app:free-chiral}  explicitly derives 
the transformation laws of the current operators in $\car$-multiplet 
in a $\cn=1$ supersymmetric theory with a free chiral multiplet on $\mathbb R\times S^3$.

\section{New minimal supergravity and Killing spinor}
\label{sec:nm-sugra}
\setcounter{equation}{0}

In this section, we briefly review the symmetries of the new-minimal supergravity and define a GKS and a Killing vector, as preliminaries for deriving the Ward identities of the $\cn=1$ field theory with an $R$-symmetry in the next section.

We begin with the construction of the new-minimal supergravity\cite{Sohnius:1981tp,Sohnius:1982fw}. It is formulated by first minimal-coupling the $\car$-multiplet 
(consisting of energy-momentum tensor $\ct^{\m\n}$, supercurrent $\cs_\a^\m$, $U(1)_R$ current $\cj^\m$ and closed two-form $\cf_{\m\n}$)
to the supergravity multiplet (containing metric $g_{\m\n}$, gravitino  $\j_{\a\m}$, $U(1)_R$ gauge field $A_\m$ and two-form gauge field $B_{\m\n}$)
to obtain the linear Lagrangian, which is  then completed to the non-linear form. 
The linear couplings take the form 
\be 
-\frac12 \ct^{\m\n}g_{\m\n}+\bar\j_\m \cs^\m+\cj^\m (A_\m-\frac32 V_\m)+\frac14\ve^{\m\n\r\l} \cf_{\m\n}B_{\r\l},
\ee 
where $V_\m=\frac14\ve_\m{}^{\n\r\l}\pa_\n B_{\r\l}$.  By definition, the vector field $V^\m$ is conserved, i.e. $\nabla_\m V^\m=0$.
In general backgrounds the operators in the $\car$-multiplet are defined in terms of the functional derivatives of the action $S$ with respect to corresponding fields in the supergravity multiplet, namely    
\begin{subequations}\label{eqn:operators}
	\begin{align}
	&\ct^{\m\n}(x)=-\frac{2}{\bm e}\frac{\d S}{\d g_{\m\n}(x)},\\
	&\cs^{\m}(x)=\frac{1}{\bm e}\frac{\d S}{\d \bar\j_{\m}(x)},\\
	& \cj^{\m}(x)=\frac{1}{\bm e}\frac{\d S}{\d A_\m(x)},\\
	& \cf_{\m\n}(x)=\frac{1}{\bm e}\ve_{\m\n\r\l}\frac{\d S}{\d B_{\r\l}(x)},
	\end{align}
\end{subequations}
where ${\bm e}\equiv|\det e^a_\m|$ with $e^a_\m$ being the vierbein.

At the classical level the new-minimal supergravity possesses the local supersymmetry as well as the $U(1)_R$ chiral symmetry and the diffeomorphism invariance. The corresponding transformation laws for the fields in the supergravity multiplet are given as follows (for conventions see Appendix \ref{app:conventions}):
\begin{itemize}
	\item Local supersymmetry transformation
\begin{subequations}		\label{eqn:nm-susy}
	\begin{align}
		& \d_\e e^a_\m=-\frac12\bar\j_\m\g^a\e,\\
		& \d_\e\j_\m=\cd_\m\e+\frac{i}{2}\g_\m(V^\r\g_\r\g_*\e),\\
		& \d_\e B_{\m\n}=\frac12(\bar\j_\m\g_\n-\bar\j_\n\g_\m)\e,\\
		&\d_\e A_\m=\frac{i}{4}(\cd_\l\bar\j_\s+\frac{i}{2}\bar\j_\s\g_*\g_\k V^\k\g_\l)\g_*\g^{\l\s}\g_\m\e.
	\end{align}
\end{subequations} 
	\item $U(1)_R$ chiral gauge transformation
\begin{subequations}		\label{eqn:nm-gauge}
	\begin{align}
		& \d_\L A_\m=\pa_\m \L,\\
		& \d_\L\j_\m=-i\g_*\j_\n\L.
	\end{align}
\end{subequations}
	\item Diffeomorphism
	\begin{subequations}\label{eqn:nm-diffeo}
		\begin{align}
			& \d_\x g_{\m\n}=\cl_\x g_{\m\n}=\nabla_\m\x_\n+\nabla_\n\x_\m,\\
			&\d_\x A_\m=\cl_\x A_\m=-F_{\m\n}\x^\n+\nabla_\m(\x^\n A_\n),\\
			&\d_\x B_{\r\l}=\cl_\x B_{\r\l}=\x^\k\nabla_\k B_{\r\l}+\nabla_\r\x^\k B_{\k\l}+\nabla_\l\x^\k B_{\r\k},\\
			& \d_\x \j_\m=\x^\n\pa_\n\j_\m+(\pa_\m\x^\n)\j_\n.
		\end{align}
	\end{subequations}
\end{itemize}
Here $F_{\m\n} \equiv \nabla_\mu A_\nu-\nabla_\nu A_\mu$.
We have omitted the higher-order terms in the gravitino, since they are irrelevant to our analysis. One can confirm the above transformation laws by checking invariance of the pure new-minimal supergravity action, given by
\be \label{eqn:nmsugra-action}
S_{\rm nm}=\frac{1}{2}\int\dd^4x\;{\bm e}\(R+6V_\m V^\m-8A_\m V^\m-\bar\j_\m\g^{\m\n\r}\cd_\n\j_\r+\text{(4 fermion terms)}\),
\ee 
where $R$ is a Ricci scalar.

As mentioned in the Introduction, a field theory with rigid supersymmetry is defined on the supersymmetric backgrounds, i.e. the ones that admit at least one solution of the GKS equation. For $\cn=1$ field theories with an $R$-symmetry the GKS equation becomes
\begin{equation}\label{eqn:GKS}
\d_\e\j_\m=\cd_\m\e+\frac{i}{2}\g_\m(V^\r\g_\r\g_*\e)=0.
\end{equation}

Now let us denote a solution of \eqref{eqn:GKS} as $\z$. As we will see in section \ref{sec:ward}, there exists a conserved supercharge corresponding to $\z$.
For the discussion of the supersymmetry algebra given in section \ref{sec:ward}, we need to define a real vector\footnote{Notice that $\bar\e\g_\m\e=0$, which follows from the property of the Majorana conjugation.} (i.e. $K^*=K$)
\begin{equation} \label{K}
K= K^\m \pa_\m  \quad \text{with} \quad K^\m= \bar\h\g^\m\z . 
\end{equation}
Here $\h\equiv i\g_*\z$ is also a GKS. By using the Fierz identity and the integrability condition for the GKS $\z$, one can derive following relations \cite{Cassani:2012xx}:
\begin{subequations}\label{eqn:killing-vector-relation}
	\begin{align}
	& K^\m\g_\m\z=0,\quad K^\m K_\m=0,\\
	&\cl_K\z=K^\m\nabla_\m\z+\frac14\nabla_\m K_\n\g^{\m\n}\z=-K^\m A_\m(i\g_*\z),\label{eqn:lie-derivative-zeta}\\
	& \nabla_\m K_\n=-\ve_{\m\n\r\s}V^\r K^\s,\quad K^\m\nabla_\m V_\n=0,\\
	& F_{\m\n}K^\m=0.
	\end{align}
\end{subequations}
It therefore follows that the background fields are invariant with respect to the null vector $K$ up to a gauge transformation for the $U(1)_R$ gauge field $A$, namely
\begin{subequations}
	\label{eqn:killing-eqns}
	\begin{align}
	& \cl_K g_{\m\n}=\nabla_\m K_\n+\nabla_\n K_\m=0,\\
	& \cl_K A_\m=-F_{\m\n}K^\n+\nabla_\m(K^\n A_\n)=\nabla_\m(K^\n A_\n),\\
	&\cl_K V_\m=-V_{\m\n}K^\n+\nabla_\m(V^\n K_\n)=0,
	\end{align}
\end{subequations}
where $V_{\m\n} \equiv\nabla_\mu V_\nu-\nabla_\nu V_\mu$.
Note that the Killing condition for $V_\m$ is not equivalent to that for $B_{\m\n}$. In fact, we do not need the Killing condition for the background field $B_{\m\n}$, as we will see in the next section.

\section{Ward identities and correlation functions  
}
\label{sec:ward}
\setcounter{equation}{0}

In this section, we derive the Ward identities of the $\cn=1$ field theory with an $R$-symmetry in 4D curved space and then comment on some properties of correlation functions, which will be basis of the discussions of the next sections.

\subsection{Ward identities}

The Ward identities corresponding to the symmetries discussed in the previous section can be obtained by using the local renormalization group formalism \cite{Osborn:1991gm} (see also \cite{Papadimitriou:2016yit} for a recent review). To this end one first defines the generating  functional of connected correlation functions
\begin{equation}
	W[g_{\m\n},\j_\m, A_\m,B_{\m\n}]= -i \log Z[g_{\m\n},\j_\m, A_\m,B_{\m\n}], 
\end{equation}
where  $Z[g_{\m\n},\j_\m, A_\m,B_{\m\n}]$ is the partition function in the presence of the non-dynamical background sources, i.e. 
$Z=\int[\cd\F]\exp{i S[\F;g_{\m\n},\j_\m, A_\m,B_{\m\n}]}$ ($\F$ represents generic matter fields), and the usual expectation values are defined as $\langle... \rangle\equiv Z^{-1}\int[\cd\F]...\exp{i S}$. 
The gravitino background $\j_\m$ is set to zero at the end of the computations, since we consider the bosonic backgrounds. Recall that the variation of the generating functional $W$ is given by
\be
\d W=\int\dd^4x\; {\bm e}\[-\frac12\braket{\ct^{\m\n}}\d g_{\m\n}+ \d \bar \j_\m \braket{\cs^\m} +\braket{\cj^\m}\d G_\m+\frac14\ve^{\m\n\r\l}\braket{\cf_{\m\n}}\d B_{\r\l}\],\label{eqn:var-W}
\ee
where $G_\m\equiv A_\m-\frac32 V_\m$. \eqref{eqn:var-W} gives the definition of the one-point functions of the operators in the presence of arbitrary sources.  
Namely, \eqref{eqn:var-W} implies that
\begin{subequations}\label{eqn:one-point-operators}
	\begin{align}
	& \braket{\ct^{\m\n}(x)}_{g_{\m\n}, \j_\m, A_\m,B_{\m\n}}=-\frac{2}{\bm e}\frac{\d W}{\d g_{\m\n}(x)},\\
	& \braket{\cs^{\m}(x)}_{g_{\m\n}, \j_\m, A_\m,B_{\m\n}}=\frac{1}{\bm e}\frac{\d W}{\d \bar\j_{\m}(x)},\\
	& \braket{\cj^{\m}(x)}_{g_{\m\n}, \j_\m, A_\m,B_{\m\n}}=\frac{1}{\bm e}\frac{\d W}{\d A_\m(x)},\\
	& \braket{\cf_{\m\n}(x)}_{g_{\m\n}, \j_\m, A_\m,B_{\m\n}}=\frac{1}{\bm e}\ve_{\m\n\r\l}\frac{\d W}{\d B_{\r\l}(x)}.
	\end{align}
\end{subequations}
The Ward identities corresponding to the symmetries \eqref{eqn:nm-susy}, \eqref{eqn:nm-gauge} and \eqref{eqn:nm-diffeo} are obtained 
by requiring $\d W=0$ up to potential quantum anomalies for the variations \eqref{eqn:nm-susy}-\eqref{eqn:nm-diffeo}, and the results are as follows:
\begin{align*} 0&=-\cd_\m\braket{\bar\cs^\m}+\frac{i}{2}\braket{\bar\cs^\m}\g_\m(V^\r\g_\r\g_*)+\frac12\bar\j_\m\g_\n\braket{\ct^{\m\n}}+\frac38\Big[(\bar\j_\m\g_\n+\bar\j_\n\g_\m) V^\n\braket{\cj^\m}\\
&\hskip1em-\bar\j_\r\g^\r V_\m\braket{\cj^\m}-\bar\j_\l\g_\r\ve^{\m\n\r\l}\nabla_\n\braket{\cj_\m}
-i \ve^{\m\r\s\n} \braket{\cj_\s} V_\r\bar\j_\m\g_*\g_\n\Big]\\
&\hskip1em
+\frac{i}{4}\cd_\l\bar\j_\s (-i \ve^{\l\s}{}_{\m\n}\g^\n+\d^\s_\m\g_*\g^\l-\d_\m^\l\g_*\g^\s)\braket{\cj^\m}+\frac14\bar\j_\m\g_\n\ve^{\m\n\r\l}\braket{\cf_{\r\l}},\numberthis\label{eqn:cls-ward-susy}\\
\ca_{\rm diffeo}&=\nabla^\m \braket{\ct_{\m\n}}-\braket{\cj^\m} G_{\m\n}-\nabla_\m \braket{\cj^\m} G_\n+\frac14\ve^{\m\k\r\l}\braket{\cf_{\m\k}}(\nabla_\n B_{\r\l}+\nabla_\r B_{\l\n}+\nabla_\l B_{\n\r}),\numberthis\label{eqn:cls-ward-diffeo}\\
\ca_{\rm chiral}&=\nabla_\m\braket{\cj^\m}. \numberthis\label{eqn:qt-ward-gauge}
\end{align*}
Here 
\begin{equation}
	G_{\m\n}\equiv \nabla_\m G_\n-\nabla_\n G_\m,\quad \cd_\n\bar\cs^\m\equiv \nabla_\n\bar\cs^\m-G_\n(i\g_*\bar\cs^\m)\,
\end{equation}
and the gravitino-dependent terms are omitted in \eqref{eqn:cls-ward-diffeo} and \eqref{eqn:qt-ward-gauge}. 
In \eqref{eqn:cls-ward-susy} it is assumed that there is no supersymmetry anomaly.

The diffeomorphism anomaly $\ca_{\rm diffeo}$ and the $U(1)_R$ chiral anomaly $\ca_{\rm chiral}$ need further explanation. The chiral anomaly is usually accompanied by the mixed gravitational anomaly, which breaks the classical diffeomorphism invariance. The anomalies are given by
\begin{align}
& \ca_{\rm chiral}=\frac14\ve^{\k\s\a\b}\[c_A G_{\k\s}G_{\a\b}+(1-\a)c_m R^\n{}_{\l\k\s}R^\l{}_{\n\a\b} \],
\label{ca}\\
& \ca_{\rm diffeo}=-\a c_m g^{\m\n}\frac{1}{\sqrt{-g}}\pa_\l\[\sqrt{-g}\frac12\ve^{\k\s\a\b}G_{\k\s}\pa_\a\G^\l_{\mu\b} \],
\end{align} 
see e.g. \cite{Jensen:2012kj} for a recent review. Here the coefficients $c_A$ and $c_m$ are determined according to the field content of the theory and are related to the central charges. The scheme parameter $\a$ is the coefficient of the diffeomorphism and gauge non-invariant contact counterterm that determines where   
the mixed anomaly appears: If $\a=0$, then the mixed anomaly appears only in the $U(1)_R$ Ward identity, while if $\a=1$ it appears only in the diffeomorphism Ward identity. In this note we choose a scheme $\a=0$ such that the mixed anomaly does not appear in the diffeomorphism Ward identity.

We emphasize that
even if we mainly consider bosonic backgrounds in quantum field theory, we must keep the gravitino background field in \eqref{eqn:cls-ward-susy} in order to compute two-point functions of the supercurrent operator. In principle, the Ward identities \eqref{eqn:cls-ward-diffeo} and \eqref{eqn:qt-ward-gauge} also contain the gravitino-dependent terms, which we ignore since they are irrelevant unless 
we differentiate \eqref{eqn:cls-ward-diffeo} and \eqref{eqn:qt-ward-gauge} with respect to the gravitino. 

\subsection{Higher-point correlation functions}

Taking further derivatives of \eqref{eqn:one-point-operators} with respect to the sources gives higher-point correlation functions, for which we use double bra-ket notation $\llangle... \rrangle$, i.e.
\be \label{correlator}
\llangle \co_{j_1}(x_1)\co_{j_2}(x_2)\cdots \co_{j_n}(x_n)\rrangle \equiv  \frac{\D }{\D \cb_{j_n}(x_n)}...\frac{\D  \, \langle \co_{j_1}(x_1) \rangle}{\D \cb_{j_2}(x_2)}
=  \frac{\D }{\D \cb_{j_n}(x_n)}...\frac{\D }{\D \cb_{j_2}(x_2)} \frac{\D \,(i W) }{\D \cb_{j_1}(x_1)}.
\ee
Here  $\co_j$  stands for any operator in the $\mathcal{R}$-multiplet, i.e. $\co_j = \left\{ \ct^{\m\n},\;\cs_\a^\m,\;\cj^\m,\; \cf^{\m\n} \right\}$, and $\frac{\D }{\D \cb_j}$ is a suitably defined functional derivative 
with respect to the background sources $\cb_j= \left\{g_{\m\n},\bar\j_{\m}, A_\m, B_{\r\l} \right\}$ as    
\be
\frac{\D }{\D \cb_j(x)} = \left\{-\frac{2}{i\bm e}\frac{\d }{\d g_{\m\n}(x)},\,\, \frac{1}{i\bm e}\frac{\d }{\d \bar\j_{\m}(x)}, \,\, \frac{1}{i\bm e}\frac{\d }{\d A_\m(x)}. \,\,
\frac{\ve_{\m\n\r\l}}{i\bm e}\frac{\d }{\d B_{\r\l}(x)}
 \right\}, 
\ee
so that $\co_j(x) = \frac{ \D (i  S)}{\D \cb_j(x)}$ according to \eqref{eqn:operators}. 
For instance,
\begin{align}
& \llangle \ct^{\m\n}(x) \cj^{\r}(y) \rrangle \equiv \frac{1}{i\bm e}\frac{\d  \braket{\ct^{\m\n}(x) }}{\d A_\r(y)} 
= \frac{1}{i\bm e}\frac{\d }{\d A_\r(y)} \left(- \frac{2}{\bm e}\frac{\d W}{\d g_{\m\n}(x)}\right) , \\
& \llangle \cs^{\m}(x) \cj^{\n}(y) \rrangle \equiv  \frac{1}{i\bm e}\frac{\d  \braket{\cs^{\m}(x) }}{\d A_\n(y)}
=   \frac{1}{i\bm e}\frac{\d   }{\d A_\n(y)} \left( \frac{1}{\bm e}\frac{\d W}{\d \bar\j_{\m}(x)} \right), \\
& \llangle \cs^{\m}(x) \cj^{\n}(y) \cj^{\r}(z) \rrangle \equiv  \frac{1}{i\bm e}\frac{\d  \llangle \cs^{\m}(x) \cj^{\n}(y) \rrangle}{\d A_\r(z)}
=   \frac{1}{i\bm e}\frac{\d   }{\d A_\r(z)} \left(  \frac{1}{i\bm e}\frac{\d   }{\d A_\n(y)} \left( \frac{1}{\bm e}\frac{\d W}{\d \bar\j_{\m}(x)} \right) \right) .
\end{align}
Notice that we use different notation $\llangle... \rrangle$ than the usual one  $\langle ... \rangle_C$ (the subscript $C$ stands for ``connected''), though $\llangle... \rrangle$  is clearly a \emph{connected} correlation function as $W$ is the generating functional for connected correlation functions. The reason of using  $\llangle ... \rrangle$ rather than $\langle ... \rangle_C$ to denote two- and higher-point functions\footnote{For  one-point functions we use the usual bra-ket notation $\langle \cdot \rangle$ since in this  case both quantities are identical, i.e. $\llangle \co(x) \rrangle = \langle  \co(x) \rangle$} is that the quantity $\llangle... \rrangle$ can differ from  $\langle... \rangle_C$ by contact terms when the operators depend on the sources due to the non-linear dependence of the action on the sources. For instance, 
using the definition of $\llangle... \rrangle$ given above it can be easily seen that
\begin{eqnarray}
&& \llangle \cj^{\m}(x) \cj^{\n}(y) \rrangle = \frac{1}{i\bm e}\frac{\d  \langle \cj^{\m}(x) \rangle }{\d A_\n(y)}
= \left \langle  \left(\frac{1}{\bm e} \frac{\d S}{\d A_\n(y)} + \frac{1}{i\bm e} \frac{\d}{ \d A_\n(y)}  \right)  \cj^{\m}(x)  \right \rangle_C  \nonumber \\
&& = \left \langle \cj^{\m}(x)  \cj^{\n}(y)  \right \rangle_C -i \delta^4(x,y) \left \langle\frac{\partial \cj^\m(x)}{\partial A_\n(x)} \right \rangle,
\end{eqnarray}      
where $\delta^4(x,y)= \delta^4(x-y)/\bm e$ is the invariant Dirac delta function. In obtaining the last term in the second line above it is used that the action (and hence operator $\cj$) does not depend on the derivatives of the background sources which are non-dynamical. 
The contact term above corresponds to the second functional derivative of action $S$ with respect to $A$, i.e. $\left \langle \frac{\d}{\bm e \d A_\n(y)}\frac{\d S}{\bm e \d A_\m(x)} \right \rangle$, which is non-vanishing 
when the action $S$ depends non-linearly on the gauge field $A$. For higher-point functions there can be more contact terms, and in general  it is not obvious whether or not the contact terms play any role and can be ignored in the calculations.  

In order to investigate the potential consequences of the contact terms, let us consider  $(n+1)$-point functions obtained 
by taking functional derivatives of an one-point function of an operator $\co(x)$  
with respect to the background sources (see \eqref{correlator}), i.e.
\begin{align}
 \llangle \co(x)\co_{j_1}(x_1)\cdots \co_{j_n}(x_n)\rrangle   & = \frac{\D }{\D \cb_{j_n}(x_n)}...\frac{\D }{\D \cb_{j_2}(x_2)} \frac{\D }{\D \cb_{j_1}(x_1)} \langle \co(x) \rangle \nonumber \\
& =  \left \langle \left(\co_{j_n}(x_n) + \frac{\D }{\D \cb_{j_n}(x_n)} \right)...\left(\co_{j_1}(x_1) + \frac{\D }{\D \cb_{j_1}(x_1)} \right) \co(x) \right \rangle_C \nonumber \\
& = \left \langle  \co_{j_n}(x_n)\cdots \co_{j_1}(x_1)  \co(x)  \right \rangle_C + \textrm{contact terms} , \label{O-n}
\end{align}
where the contact terms contain connected correlation functions with less than $n+1$ operators and are proportional to $\delta$-functions that arise whenever $ \frac{\D }{\D \cb}$ acts 
on operators.  The contact terms can be split into two parts according to whether it contains $\delta^{4}(x,x_k)$ or not, namely      
\be \label{split-contact}
\textrm{contact terms} =\quad  \textrm{terms with } \delta^{4}(x,x_k)  \quad + \quad \langle \co(x)... \rangle_C  \, ,
\ee
where the first part on the RHS of \eqref{split-contact} consists of terms that contain $\delta^{4}(x,x_k)$ (and their products), which results from $ \frac{\D }{\D \cb_{j_k}}$ acting on $\co(x)$, while the rest is collected into the second term, which  does not contain the derivatives of $\co(x)$ and hence can be written in the form of $\langle \co(x)... \rangle_C$ 
The second part in \eqref{split-contact} contains $\delta^4(x_k,x_l)$ (and their products) with $k\neq l$ and is absent when $n=1$ (i.e. for two-point functions). Splitting in the form \eqref{split-contact} will be useful in the following discussions.  

Now we suppose that
$\co(x)$ corresponds to a conserved current $X^\m$, i.e. $\co(x)=X^\m(x)$ with $\nabla_\mu \langle X^\m(x) \rangle=0$. When acting 
$\nabla_\mu = \bm e^{-1} \frac{\partial}{\partial x^\mu} \bm e$ on \eqref{split-contact} and taking integration $\int\dd^4x\;{\bm e}$, the first part in \eqref{split-contact} does not contribute since it leads to the integration of the total derivative of $\delta^4(x-x_j)$ that vanishes. In the case of two-point functions, the second part in \eqref{split-contact} does not exist. Therefore, for the two-point functions all contact terms drop out through the operations mentioned above, so that we have 
\be \label{2pt-X-Q}
 \int\dd^4x\;{\bm e} \nabla_\mu \llangle X^\m(x) \co_{j_1}(y)\rrangle= \int\dd^4x\;{\bm e} \nabla_\mu \langle X^\m(x) \co_{j_1}(y)\rangle_C
 =  \langle [Q_X, \co_{j_1}(y)]\rangle ,
\ee 
where $Q_X$ is the corresponding conserved charge defined as $Q_X \equiv \int_\cc\dd\s_\m\;X^\m$ with $\cc$ 
being the Cauchy surface.\footnote{One comment is in order on the second equality of \eqref{2pt-X-Q}. The usual relation is $\langle [Q_X, \co_j(y)]\rangle= \int\dd^4x\;{\bm e} \nabla_\mu \langle X^\m(x) \co_j(y)\rangle$ without subscript $C$. However, since $\nabla_\mu \langle X^\m(x) \rangle=0$, the disconnected connected part vanishes, so that $\langle [Q_X, \co_j(y)]\rangle= \int\dd^4x\;{\bm e} \nabla_\mu \langle X^\m(x) \co_j(y)\rangle_C$.} 
\eqref{2pt-X-Q} will be often used in section \ref{SUSY}. For three-point and higher-point functions the second part of the contact terms in \eqref{split-contact} can contribute and needs a careful treatment. 

Let us consider the case when $Q_X$ annihilates the vacuum state $\ket{\O}$, i.e. $Q_X  \ket{\Omega}=0$. In this case it follows that 
\be \label{npt-X-Q}
Q_X  \ket{\Omega}=0
\,\,\implies
 \,\, \int\dd^4x\;{\bm e} \nabla_\mu \llangle X^\m(x) \co_{j_1}(x_1)\cdots \co_{j_n}(x_n)\rrangle =0, 
\ee 
This is analogous to the usual formula  
$\int\dd^4x\;{\bm e} \nabla_\mu \langle X^\m(x) ...\rangle =0$ when $Q_X  \ket{\Omega}=0$. 
 \eqref{npt-X-Q} can be shown as follows. First, note that $Q_X \ket{ \Omega}=0$ leads to $\int\dd^4x\;{\bm e} \nabla_\mu \langle X^\m(x)...\rangle_C=0$, which follows 
from $\int\dd^4x\;{\bm e} \nabla_\mu \langle X^\m(x) ...\rangle =0$ and the definition of the connected correlation functions.\footnote{As an illustration of the statement that $\int\dd^4x\;{\bm e} \nabla_\mu \langle X^\m(x)...\rangle_C=0$ 
if $\nabla_\mu \langle X^\m(x) ...\rangle=0$,  
let us consider connected three-point function  
$ \int\dd^4x\;{\bm e} \nabla_\mu \langle X^\m(x) \co_{j_1}(x_1) \co_{j_2}(x_2)\rangle_C$. From the  definition of the connected correlation function we have 
$ \int\dd^4x\;{\bm e} \nabla_\mu \langle X^\m(x) \co_{j_1}(x_1) \co_{j_2}(x_2)\rangle_C= \int\dd^4x\;{\bm e} \nabla_\mu \langle X^\m(x) \co_{j_1}(x_1) \co_{j_2}(x_2)\rangle - 
\int\dd^4x\;{\bm e} \nabla_\mu \langle X^\m(x) \rangle\langle \co_{j_1}(x_1) \co_{j_2}(x_2)\rangle-
\langle \co_{j_1}(x_1) \rangle \int\dd^4x\;{\bm e} \nabla_\mu \langle X^\m(x)  \co_{j_2}(x_2)\rangle-
 \int\dd^4x\;{\bm e} \nabla_\mu \langle X^\m(x)  \co_{j_1}(x_1)\rangle \langle \co_{j_2}(x_1) \rangle 
 + 2 \int\dd^4x\;{\bm e} \nabla_\mu \langle X^\m(x) \rangle\langle \co_{j_1}(x_1) \rangle\langle \co_{j_2}(x_2)\rangle
$, each term of which contains 
$\int\dd^4x\;{\bm e} \nabla_\mu \langle X^\m(x)...\rangle=0$, 
so  
we have  
$\int\dd^4x\;{\bm e} \nabla_\mu \langle X^\m(x) \co_{j_1}(x_1) \co_{j_2}(x_2)\rangle_C=0$. This fact is easily generalized to the arbitrary higher-point functions.   
}
Then, the first term (non-contact part) in \eqref{O-n} (with $\co(x)=X^\m(x)$) and the second part of the contact terms 
in \eqref{split-contact} do not contribute to $\int\dd^4x\;{\bm e} \nabla_\mu \llangle X^\m(x) \co_{j_1}(x_1)\cdots \co_{j_n}(x_n)\rrangle$. Since  
the first part of the contact terms in \eqref{split-contact} does not contribute either as mentioned above, we end up with \eqref{npt-X-Q}, which will be employed  in section \ref{sec:quantum-consistency}.

\section{Rigid supersymmetry algebra} \label{SUSY}

Now we recover the rigid supersymmetry algebra on the curved backgrounds by deriving the transformation laws of the supercurrent and $U(1)_R$ current with respect to the rigid supersymmetry.\footnote{We suspect that these transformation rules should be known, but we found only their flat-space version in the literature, see e.g. \cite{Sohnius:1981tp,Closset:2013vra}.} 

In order to set up the general strategy, we first deal with the diffeomorphism Ward identity. 
We multiply \eqref{eqn:cls-ward-diffeo} by an arbitrary vector field $\x^\n(x)$ and take a functional derivative  $\frac{\delta }{i{\bm e}\, \delta A_\rho(y)}$ to obtain
\bala
0&=i\x^\n\Big[\nabla^\m\braket{\braket{\ct_{\m\n}(x)\cj^\r(y)}}-G_{\m\n}\braket{\braket{\cj^\m(x)\cj^\r(y)}}-G_\n\nabla_\m\braket{\braket{\cj^\m(x)\cj^\r(y)}}\\
&\hskip1em+\frac14\ve^{\m\k\r\l}(\nabla_\n B_{\r\l}+\nabla_\r B_{\l\n}+\nabla_\l B_{\n\r})\braket{\braket{\cf_{\m\k}(x)\cj^\r(y)}}\Big]\\
&\hskip1em-\d^4(x,y)[\nabla_\n(\x^\n\braket{\cj^\r})-\braket{\cj^\n}\nabla_\n \x^\r]
+e^{-1}\pa_\n[\d^4(x-y)(\x^\n\braket{\cj^\r}-\x^\r\braket{\cj^\n})].\numberthis\label{eqn:stress-R-two-point}
\eala
The integration over $x$-space (i.e. $\int\dd^4x\;{\bm e}$) of the above gives
\begin{align}
	&\nabla_\n(\x^\n\braket{\cj^\r})-\braket{\cj^\n}\nabla_\n \x^\r=\int\dd^4x\;{\bm e}\;i\x^\n\Big[\nabla^\m\braket{\braket{\ct_{\m\n}(x)\cj^\r(y)}}-
	G_{\m\n}\braket{\braket{\cj^\m(x)\cj^\r(y)}}-\NO \\
	&\hskip3em-G_\n\nabla_\m\braket{\braket{\cj^\m(x)\cj^\r(y)}}+\frac14\ve^{\m\k\r\l}(\nabla_\n B_{\r\l}+\nabla_\r B_{\l\n}+
	\nabla_\l B_{\n\r})\braket{\braket{\cf_{\m\k}(x)\cj^\r(y)}}\Big],\label{eqn:diffeo-var-R}
\end{align}
where the left-hand side 
actually corresponds to the variation of the operator $\cj^\r$ under the diffeomorphism associated with the vector $\x^\m$, see e.g. section 5.2.3 in \cite{DiFrancesco:1997nk}. It then follows from \eqref{eqn:diffeo-var-R} that the $U(1)_R$ current $\cj^\m$ transforms as a vector density under the diffeomorphism, which 
has to do with the fact that 
the quantity conjugate to the vector source is not a vector but a vector density operator (see e.g. \cite{Papadimitriou:2010as}).

Multiplying \eqref{eqn:cls-ward-diffeo} by $K^\n$ defined in \eqref{K} and assuming that $K^\m A_\m$ is made constant\footnote{When $K^\m A_\m$ is not constant, it  seems that there are several inconsistencies in constructing the supersymmetry algebra. For instance, one can not define a conserved charge associated with the Killing vector $K$, and the Lie derivative of $\z$ with respect to $K$ does not satisfy the generalized Killing condition.} by a suitable $U(1)_R$ gauge transformation,
we obtain\footnote{A careful derivation is necessary for the term containing $\cf_{\r\l}$. First of all, we have $\nabla_\m(\ve^{\m\n\r\l}K_\n\cf_{\r\l})=4V^\m K^\n\cf_{\m\n}=\ve^{\m\r\s\l}K^\n(\nabla_\r B_{\s\l})\cf_{\m\n}$. Combining this with $\ve^{[\m\r\s\l}K^{\n]}(\nabla_\r B_{\s\l})\cf_{\m\n}=0$, which implies $\ve^{\m\r\s\l}K^\n(\nabla_\r B_{\s\l})\cf_{\m\n}=\frac12\cf_{\m\n}\ve^{\s\l\m\n}K^\r(\nabla_\r B_{\s\l}+\nabla_\l B_{\r\s}+\nabla_\s B_{\l\r})$, gives 
$\nabla_\m(\ve^{\m\n\r\l}K_\n\cf_{\r\l})=\frac12\cf_{\m\n}\ve^{\s\l\m\n}K^\r(\nabla_\r B_{\s\l}+\nabla_\l B_{\r\s}+\nabla_\s B_{\l\r})$. Notice that we do not need the Killing condition for $B_{\m\n}$.}
\begin{equation} \label{cc_K}
\nabla_\m \braket{\cc_K^\m(x)}=0,
\end{equation}
where
\begin{equation} \label{cc_K_def}
	\cc^\m_K(x)\equiv K_\n\Big[\ct^{\m\n}-\cj^\m(A^\n-\frac32 V^\n)+\frac12\ve^{\m\n\r\l}\cf_{\r\l} \Big](x).
\end{equation}
This allows us to define a conserved charge
\begin{equation}
	Q_K\equiv \int_\cc\dd\s_\m\;\cc^\m_K,
\end{equation}
where $\cc$ is any Cauchy surface. Now using \eqref{2pt-X-Q} and replacing $\x$ by $K$ in \eqref{eqn:diffeo-var-R}, we obtain
\begin{equation}\label{eqn:diffeo-K-J}
	i\langle [Q_K,\cj^\r]\rangle=\int\dd^4x\;{\bm e}\;i\nabla_\m\llangle\cc^\m_K(x)\cj^\r(y)\rrangle=\nabla_\n(K^\n \langle\cj^\r\rangle)-\langle\cj^\n\rangle \nabla_\n K^\r.
\end{equation}

Now we use the above strategy to recover the rigid supersymmetry algebra for $\cn=1$ field theories with an $R$-symmetry. 
Multiplying \eqref{eqn:cls-ward-susy} by the GKS $\z(x)$ and setting the gravitino to zero gives 
\begin{equation}
\nabla_\m\left( \langle \bar\cs^\m \z(x) \rangle \right) =0 ,
\end{equation}
which allows us to define a conserved supercharge associated with the GKS $\z$ as 
\begin{equation}
\cq_\z\equiv\int_\cc \dd\s_\m\;\bar\cs^\m\z .
\end{equation}
Independence of $\cq_\z$ on the choice for the Cauchy surface, i.e. conservation of $\cq_\z$ is an immediate consequence of the Ward identity \eqref{eqn:cls-ward-susy} on the bosonic background. Note that we can also define a conserved supercharge $\cq_\h$ associated with the GKS $\h=i\g_*\z$.
Now we  multiply \eqref{eqn:cls-ward-susy} by the GKS $\z$ and  taking the functional derivative  $\frac{1}{i\bm e} \frac{\delta}{\delta\bar\j_\m(y)}$
 to obtain
\begin{align*}
	i\nabla_\n\llangle(\bar\cs^\n\z)(x)\cs^\m(y)\rrangle &=\d^4(x,y)\Big[\frac14\g_\n\z\Big(2\braket{\ct^{\m\n}}+3\braket{\cj^\m} V^\n
	+\braket{\cj^\n} V^\m-g^{\m\n}\braket{\cj^\r} V_\r+\ve^{\m\n\r\l}\braket{\cf_{\r\l}}\\
	&\hskip1em  -\frac12 \ve^{\m\n\l\s}\nabla_\l\braket{\cj_\s}\Big)+\frac{i}{4}\g_*\g_\n\z(g^{\m\n}\nabla_\r\braket{\cj^\r}-\nabla^\n\braket{\cj^\m}+\ve^{\m\n\s\k}\braket{\cj_\s} V_\k)\Big]\\
	&\hskip1em+\frac{i}{4} \bm e^{-1}\pa_\l\big[\d^4(x-y)(-i \ve^{\l\m}{}_{\s\n}\g^\n+\d^\m_\s\g_*\g^\l-\d_\s^\l\g_*\g^\m)\z\braket{\cj^\s}\big],\numberthis\label{eqn:supercurrent-two-point}
\end{align*}
where we 
have set the gravitino background to zero at the end.  
Integrating \eqref{eqn:supercurrent-two-point} over $x$-space and using \eqref{2pt-X-Q} give
\bala 
& i\langle [\cq_\z,\cs^\m]\rangle=\int\dd^4x\;{\bm e}\;i\nabla_\n\llangle (\bar\cs^\n\z)(x)\cs^\m(y)\rrangle\\
&=\frac14\g_\n\z\left(2\braket{\ct^{\m\n}}+3\braket{\cj^\m }V^\n+\braket{\cj^\n} V^\m-g^{\m\n}\braket{\cj^\r} V_\r+\ve^{\m\n\r\l}\braket{\cf_{\r\l}}
-\frac12 \ve^{\m\n\l\s}\nabla_\l\braket{\cj_\s}\right)+\\
&
\quad +\frac{i}{4}\g_*\g_\n\z\big(g^{\m\n}\nabla_\r\braket{\cj^\r}-\nabla^\n\braket{\cj^\m}+\ve^{\m\n\s\k}\braket{\cj_\s} V_\k\big). \numberthis\label{eqn:supercurrent-transformation}
\eala 
Note that using \eqref{eqn:supercurrent-transformation} we can rewrite \eqref{eqn:supercurrent-two-point} in a simple form as 
\begin{equation}
	\nabla_\n\braket{\braket{(\bar\cs^\n\z)(x)\cs^\m(y)}}=
	\d^4(x,y)\braket{[\cq_\z,\cs^\m]}+\frac{1}{4}\bm e^{-1}\pa_\l\big[\d^4(x-y)(-i \ve^{\l\m}{}_{\s\n}\g^\n+\d^\m_\s\g_*\g^\l-\d_\s^\l\g_*\g^\m)\z\braket{\cj^\s}\big].\label{eqn:susy-s-s}
\end{equation}
Multiplying \eqref{eqn:supercurrent-transformation} by $\bar\z$ and $\bar\h$ gives respectively 
(omitting the bra-ket notation $\braket{\cdot}$)
\begin{align}
& i[\cq_\z,\bar\z\cs^\m]=\frac14\nabla_\r(K^\m\cj^\r-K^\r\cj^\m),\label{eqn:Qzeta-zetaS}\\
& i[\cq_\z,\bar\h\cs^\m]=\frac12K_\n[\ct^{\m\n}-G^\n\cj^\m+\frac12\ve^{\m\n\r\l}\cf_{\r\l}]-\frac18\nabla_\l(\ve^{\m\n\l\s}K_\n\cj_\s)+\frac12(K^\n A_\n)\cj^\m.\label{eqn:Qzeta-etaS}
\end{align}
It then follows that
\begin{align}
&	\nabla_\m[\cq_\z,\bar\z\cs^\m]=0,\quad  [\cq_\z,\cq_\z]=0,\\
&	i[\cq_\z,\cq_\h]=\frac12Q_K+\frac12(K^\n A_\n)Q_R,
\end{align}
where $Q_R$ is the $U(1)_R$ charge.

Now we multiply \eqref{eqn:cls-ward-susy} by the GKS $\z(x)$, differentiate it  with respect to $A_\m(y)$ and $B_{\s\l}(y)$, respectively (cf. \eqref{eqn:one-point-operators}), and set the gravitino to zero. Then, we get
\be 
\nabla_\n\llangle(\bar\cs^\n\z)\cj^\m\rrangle= \braket{\bar\cs^\m}\g_*\z\d^4(x,y),\quad \nabla_\r\llangle(\bar\cs^\r\z)\cf_{\m\n}\rrangle=
\frac{1}{2}\Big[\nabla_\m(\braket{\bar\cs^\r}\g_\r\g_\n\g_*\z)-(\m\leftrightarrow\n)\Big]\d^4(x,y),
\ee 
which lead to 
\begin{align}
&[\cq_\z,\cj^\m]=\bar\cs^\m\g_*\z,\quad i[\cq_\z,Q_R]=\cq_\h,\label{eqn:susy-var-j-main}\\
&[\cq_\z,\cf_{\m\n}]=\frac{1}{2}\Big[\nabla_\m(\bar\cs^\r\g_\r\g_\n\g_*\z)-(\m\leftrightarrow\n)\Big].\label{eqn:susy-var-cf-main}
\end{align}
These transformation laws of the currents $\cs^\m$, $\cj^\m$ and $\cf_{\m\n}$ under the rigid supersymmetry are explicitly checked in appendix \ref{app:free-chiral} for a free chiral theory on $\mathbb R\times S^3$. 

Transformation law for the supercharge $\cq_\z$ under the diffeomorphism associated with the Killing vector $K$ can be obtained by differentiating the diffeomorphism Ward identity with respect to the gravitino source. For this, the diffeomorphism Ward identity should be extended to involve the gravitino-dependent terms. We do not present the details of its calculation here, but give the final result as
\begin{equation}
	i[Q_K,\cq_\z]=-(K^\m A_\m)\cq_\h \quad \text{or}\quad [Q_K+(K^\m A_\m)Q_R,\cq_\z]=0,
\end{equation}
which is consistent with \eqref{eqn:lie-derivative-zeta}.

In summary, the  
supersymmetry algebra is
\begin{subequations}\label{eqn:susyalgebra}
	\begin{align}
	& i[\cq_\z,\cq_\h]=\frac12Q_K+\frac12(K^\n A_\n)Q_R,\label{eqn:susyalgebra-zeta-eta}\\
	& [Q_K+(K^\m A_\m)Q_R,\cq_\z]=0,\quad i[\cq_\z,Q_R]=\cq_\h,\\
	&[Q_K+(K^\m A_\m)Q_R,\cq_\z]=0,
	\end{align}
\end{subequations}
see e.g. \cite{Dumitrescu:2012ha}.

We end this section by addressing  the quantum consistency of the $\cn=1$ rigid supersymmetry algebra \eqref{eqn:susyalgebra}. First, note that by differentiating \eqref{eqn:qt-ward-gauge} with respect to the source field $A_\n(y)$, we get
\begin{equation}
	i\nabla_\m\braket{\braket{\cj^\m(x)\cj^\n(y)}}=c_A\ve^{\m\n\r\s}\nabla_\m\d^4(x,y)G_{\r\s},
\end{equation}
where the right-hand side is a total derivative. It then follows that 
\begin{equation}
	\int\dd^4x\;{\bm e}\;\nabla_\m\braket{\braket{\cj^\m(x)\cj^\n(y)}}=0,
\end{equation} 
which implies that $[Q_R,\cj^\n]=0$. Using this, we find from \eqref{eqn:susyalgebra-zeta-eta} and \eqref{eqn:diffeo-K-J} that
\begin{equation}
	2[[\cq_\z,\cq_\h],\cj^\m]=-i[Q_K,\cj^\m]=-\nabla_\n(K^\n\cj^\m)+\cj^\n\nabla_\n K^\m.\label{eqn:zeta-eta-J-1}
\end{equation}
On the other hand, \eqref{eqn:susy-var-j-main} and \eqref{eqn:Qzeta-zetaS} imply that
\begin{equation}
	2[\cq_\z,[\cq_\h,\cj^\m]]=2i[\cq_\z,\bar\z\cs^\m]=\frac12\nabla_\n(K^\m\cj^\n-K^\n\cj^\m),
\end{equation}
and therefore
\begin{equation}
	2[[\cq_\z,\cq_\h],\cj^\m]=2[\cq_\z,[\cq_\h,\cj^\m]]-2[\cq_\h,[\cq_\z,\cj^\m]]=\nabla_\n(K^\m\cj^\n-K^\n\cj^\m).\label{eqn:zeta-eta-J-2}
\end{equation}
Since $K$ is a nowhere vanishing vector \cite{Dumitrescu:2012ha}, \eqref{eqn:zeta-eta-J-1} can be consistent with \eqref{eqn:zeta-eta-J-2} only when
\begin{equation}
\nabla_\m\braket{\cj^\m}=0,  \label{eqn:new-constraint-J}
\end{equation}
or equivalently (see \eqref{eqn:qt-ward-gauge} and \eqref{ca})
\begin{equation}
	\ve^{\m\n\r\s}G_{\m\n}G_{\r\s}=0, \quad \ve^{\k\s\a\b}R^\n{}_{\l\k\s}R^\l{}_{\n\a\b}=0, \label{eqn:new-constraint}
\end{equation}
if we assume $c_A\neq 0$ and $c_m\neq0$. These are 
additional constraints imposed on the background sources since the GKS condition \eqref{eqn:GKS} does not automatically imply \eqref{eqn:new-constraint} \cite{Cassani:2013dba}. Does the condition \eqref{eqn:new-constraint} suffice  
for consistent construction of $\cn=1$ field theory with an $R$-symmetry when  $c_A\neq 0$ and $c_m\neq0$? As we will see in the next section the answer is no,  
due to a problem that manifests itself in the higher-point correlation functions. 

\section{Quantum consistency 
}
\label{sec:quantum-consistency}
\setcounter{equation}{0}

In this section we show that the $U(1)_R$ coefficient $c_A$ 
should vanish in order for the new-minimal supergravity formulation of the $\cn=1$ field theory with an $R$-symmetry to be quantum-mechanically consistent.

Condition \eqref{eqn:new-constraint}  is related to the $\cn=1$ rigid supersymmetry algebra \eqref{eqn:susyalgebra}, which relies on the assumption that $K^\m A_\m$ is constant. 
Now we would like to 
pursue our investigation  
without using this assumption. 
Instead, following \cite{Assel:2015nca} we suppose that the vacuum state $\ket{\O}$ 
is supersymmetric, i.e.
\begin{equation}
	\cq_\z\ket{\O}=\cq_\h\ket{\O}=0, \label{Q_z_h}
\end{equation}
which implies $\langle\delta_\z(...) \rangle=\langle\delta_\h(...) \rangle=0$ with $\delta_\z(...)\equiv [\cq_\z , \, ...]$ and $\delta_\h(...)\equiv [\cq_\h , \,...]$.  
Notice that when $ K^\m A_\m $ is not constant, the right-hand side of \eqref{cc_K} becomes non-zero, i.e.
$ \nabla_\m \cc^\m=-\cj^\m \nabla_\m (A_\n K^\n)$,
and \eqref{eqn:Qzeta-etaS} therefore leads to 
\begin{equation}
0=i\nabla_\m\braket{[\cq_\z,\bar\h\cs^\m]}=\frac12 K^\n A_\n\nabla_\m\braket{\cj^\m}= \frac12 K^\n A_\n
\ca_{\rm chiral} \label{eqn:one-point-inconsistency}
\end{equation}
on the supersymmetric vacuum. This implies that
\begin{equation}
K^\n A_\n=0 \quad \text{or}\quad \nabla_\m\braket{\cj^\m}=\ca_{\rm chiral}=0.
\end{equation}
Yet this does not seem to cause a serious problem, as we saw the similar constraints 
in the previous section, see \eqref{eqn:new-constraint-J}-\eqref{eqn:new-constraint}.
However, it turns out that the real problem shows up in the higher-point functions.
To see this, we multiply \eqref{eqn:cls-ward-susy} by any spinor $\e(x)$ and differentiate with respect to $\bar \j_\m(x_1)$ and then $A_\n(x_2)$. We then set the gravitino to zero and obtain
\bala 
& i\cd_\r\llangle \bar\cs^\r\e(x)\cs^\m(x_1)\cj^\n(x_2)\rrangle +\frac{1}{2}\llangle\bar\cs^\l\g_\l V^\r\g_{\r}\g_*\e(x)\cs^\m(x_1)\cj^\n(x_2)\rrangle=\\
&=\frac12\d^4(x_1,x)\g_\l\e\Big \langle \Big \langle \Big[\ct^{\m\l}+\frac34(V^\l\cj^\m+V^\m\cj^\l-g^{\m\l}V_\r \cj^\r-\ve^{\m\l\r\k}\nabla_\r\cj_\k)+\frac12\ve^{\m\l\r\k}\cf_{\r\k}\Big](x_1)\cj^\n(x_2)
\Big \rangle \Big \rangle\\
&\hskip1em+\frac{3i}{8}\d^4(x_1,x)\g_*\g_\l\e\;\ve^{\m\l\r\s}V_\s \llangle \cj_\r(x_1) \cj^\n(x_2)\rrangle  \\
&\hskip1em-\frac{i}{4} \d^4(x_1,x)(-i \ve^{\l\m}{}_{\s\k}\g^\k+\d^\m_\s\g_*\g^\l-\d_\s^\l\g_*\g^\m)\cd_\l\big(\e\llangle \cj^\s(x_1)\cj^\n(x_2) \rrangle \big)\\
&\hskip1em +\frac{i}{4} \bm e^{-1}\pa_\l\big[\d^4(x_1-x)(-i \ve^{\l\m}{}_{\s\k}\g^\k+\d^\m_\s\g_*\g^\l-\d_\s^\l\g_*\g^\m)\e \llangle \cj^\s(x_1)\cj^\n(x_2)\rrangle \big]\\
&\hskip1em+\frac14\d^4(x_1,x)\d^4(x_1,x_2)(-i \ve^{\n\m}{}_{\s\k}\g^\k+\d^\m_\s\g_*\g^\n-\d_\s^\n\g_*\g^\m)\g_*\e\;\braket{\cj^\s}
+i\d^4(x_2,x)\llangle\bar\cs^\n\g_*\e(x)\;\cs^\m(x_1)\rrangle.\numberthis\label{eqn:3point-general}
\eala 
As mentioned before, it is important to keep all contact terms in the above computation. 
Now we let $\e(x)$ be the GKS $\z$  and multiply \eqref{eqn:3point-general} by $ \bar \eta= i\bar\z(x_1)\g_*$. We then obtain
\bala
&i\nabla_\r\braket{\braket{\bar\cs^\r\z(x)\;\bar\h\cs^\m(x_1)\;\cj^\n(x_2)}}= \\
&=\frac12\d^4(x_1,x) \Big\langle \Big \langle \Big[K_\k(\ct^{\m\k}-G^\k\cj^\m+\frac12\ve^{\m\k\r\l}\cf_{\r\l})-\frac14\nabla_\l(\ve^{\m\k\l\s}K_\k\cj_\s)+(K^\r A_\r)\cj^\m\Big](x_1)\cj^\n(x_2)
\Big\rangle \Big \rangle \\
&\quad +i\d^4(x_2,x)\braket{\braket{\bar\cs^\n\g_*\z(x_2)\;\bar\h\cs^\m(x_1)}}+\frac{i}{4}\d^4(x_1,x)\d^4(x_1,x_2)(K^\n\braket{\cj^\m}-K^\m\braket{\cj^\n}).\numberthis
\eala 
Let us integrate this over $x$-space, using (see \eqref{npt-X-Q})
\begin{equation}
0=\Braket{\d_\z\(\bar\h\cs^\m(x_1)\;\cj^\n(x_2)\)} \,\,\implies \,\,0=\int\dd^4x\;{\bm e}\;\nabla_\r\llangle \bar\cs^\r\z(x)\;\bar\h\cs^\m(x_1)\;\cj^\n(x_2)\rrangle,
\end{equation} 
and take a covariant divergence with respect to $x_1$ to obtain
\begin{align}
	0&=\nabla_\m \int\dd^4x\;{\bm e}\;i\nabla_\r\llangle\bar\cs^\r\z(x)\;\bar\h\cs^\m(x_1)\;\cj^\n(x_2)\rrangle 
	\NO\\
	&=\frac12(K^\l A_\l )\nabla_\m\braket{\braket{\cj^\m(x_1)\cj^\n(x_2)}}-\frac{i}{2}\d^4(x_1,x_2)K^\n\nabla_\r\braket{\cj^\r(x_2)}\NO\\
	&=\frac{K^\l}{2} \frac{\d}{i\bm e\, \d A_\n(x_2)}\Big( A_\l\nabla_\m\braket{\cj^\m} \Big).  \label{two-point-consistency}
\end{align}
For the second equality we used the relation \eqref{eqn:susy-s-s} (where the first term on the RHS vanishes due to the assumption of the supersymmetric vacuum) and
\begin{align}
0&=i\nabla_\m\Big[K_\n\Big(\braket{\braket{\ct^{\m\n}(x_1)\cj^\r(x_2)}}+\frac32\braket{\braket{\cj^\m(x_1)\cj^\r(x_2)}} V^\n
+\frac12\ve^{\m\n\r\l}\braket{\braket{\cf_{\r\l}(x_1)\cj^\r(x_2)}}\Big)\Big]\NO\\
&\hskip3em -iK^\n A_\n\nabla_\m\braket{\braket{\cj^\m(x_1)\cj^\r(x_2)}}-\d^4(x_1,x_2)\Big[\nabla_\n(K^\n\braket{\cj^\r})-\braket{\cj^\n}\nabla_\n K^\r\Big]
\NO\\
&\hskip3em +\bm e^{-1}\pa_\n\Big[\d^4(x_1-x_2)(K^\n\braket{\cj^\r}-K^\r\braket{\cj^\n})\Big],\label{eqn:diffeo-T-J}
\end{align} 
which is obtained from \eqref{eqn:stress-R-two-point} by replacing $\x^\n$ by the Killing vector $K^\n$ and using the Killing equations \eqref{eqn:killing-eqns}. Note that in the above computation there occurs a complete cancellation between the contact terms. Using \eqref{eqn:qt-ward-gauge}, \eqref{two-point-consistency} implies 
\begin{equation}
K^\l \frac{\d}{i\bm e\, \d A_\n(x_2)}\Big( A_\l\ca_{\rm chiral}(x_1) \Big)=0 , \label{eqn:two-point-consistency}
\end{equation}
which is another constraint in addition to \eqref{eqn:one-point-inconsistency}.

The analysis up to now is insufficient to say about the quantum inconsistency with the known $U(1)_R$ anomaly, 
because there may exist very restrictive backgrounds on which  the constraints \eqref{eqn:one-point-inconsistency} and \eqref{eqn:two-point-consistency}  are satisfied. 
Therefore we need to go further to higher-point functions.  
To this end, we differentiate \eqref{eqn:3point-general} once more with respect to the gauge field source
$A_\l(x_3)$ to obtain
\bala 
&i\cd_\r\llangle\bar\cs^\r\e(x)\cs^\m(x_1)\cj^\n(x_2)\cj^\l(x_3)\rrangle +\frac{1}{2}\llangle\bar\cs^\k\g_\k V^\r\g_\r\g_*\e(x)\cs^\m(x_1)\cj^\n(x_2)\cj^\l(x_3)\rrangle=\\
&=\frac12\d^4(x,x_1)\g_\r\e\Big\langle \Big \langle\Big[\ct^{\m\r}+\frac34(V^\r\cj^\m+V^\m\cj^\r-g^{\m\r}V_\k \cj^\k-\ve^{\m\r\s\k}\nabla_\s\cj_\k)+ \NO \\ 
& \hspace{7.5cm}+\frac12\ve^{\m\r\s\k}\cf_{\s\k}\Big](x_1)\cj^\n(x_2)\cj^\l(x_3)
\Big\rangle \Big \rangle\\
&\hskip1em+i\d^4(x,x_2)\llangle\bar\cs^\n\g_*\e(x)\;\cs^\m(x_1)\cj^\l(x_3)\rrangle+i\d^4(x,x_3)\llangle\bar\cs^\l\g_*\e(x)\;\cs^\m(x_1)\cj^\n(x_2)\rrangle \\
&\hskip1em +\frac{i}{4}\bm e^{-1}\pa_\r\big[\d^4(x-x_1)(-i \ve^{\r\m}{}_{\s\k}\g^\k+\d^\m_\s\g_*\g^\r-\d_\s^\r\g_*\g^\m)\e\llangle\cj^\s(x_1)\cj^\n(x_2)\cj^\l(x_3)\rangle\big]\\
&\hskip1em-\d^4(x,x_1)\frac{i}{4}(-i \ve^{\r\m}{}_{\s\k}\g^\k+\d^\m_\s\g_*\g^\r-\d_\s^\r\g_*\g^\m)\cd_\r\big(\e\llangle\cj^\s(x_1)\cj^\n(x_2)\cj^\l(x_3)\rrangle\big)\\
&\hskip1em+\frac{3i}{8}\d^4(x,x_1)\g_*\g_\r\e\;\ve^{\m\r\k\s}V_\s \llangle \cj_\k (x_1)\cj^\n(x_2)\cj^\l(x_3)\rrangle \\
&\hskip1em-\frac{i}{4}\d^4(x,x_1)\d^4(x,x_2)(-i \ve^{\n\m}{}_{\s\k}\g^\k+\d^\m_\s\g_*\g^\n-\d_\s^\n\g_*\g^\m)\g_*\e\;\llangle\cj^\s(x)\cj^\l(x_3)\rrangle\\
&\hskip1em-\frac{i}{4}\d^4(x,x_1)\d^4(x,x_3)(-i \ve^{\l\m}{}_{\s\k}\g^\k+\d^\m_\s\g_*\g^\l-\d_\s^\l\g_*\g^\m)\g_*\e\;\llangle\cj^\s(x)\cj^\n(x_2)\rrangle .\numberthis\label{eqn:4point-general}
\eala 
By following essentially the same steps as done to reach 
\eqref{eqn:two-point-consistency} from \eqref{eqn:3point-general}
for the 3-point function $\llangle \bar\cs^\r\e(x)\cs^\m(x_1)\cj^\n(x_2)\rrangle$, one can obtain
\bala
0&=\nabla_\m \int\dd^4x\;{\bm e}\;i\nabla_\r \llangle\bar\cs^\r\z(x)\;\bar\h\cs^\m(x_1)\cj^\n(x_2)\cj^\l(x_3) \rrangle 
\\
&=\frac12(K^\r A_\r)\nabla_\m\braket{\braket{\cj^\m(x_1)\cj^\n(x_2)\cj^\l(x_3)}}-\frac{i}{2}\d^4(x_1,x_2)K^\n\nabla_\m\braket{\braket{\cj^\m(x_1)\cj^\l(x_3)}}\\
& \quad 
-\frac{i}{2}\d^4(x_1,x_3)K^\l\nabla_\m\braket{\braket{\cj^\m(x_1)\cj^\n(x_2)}}\\
&=\frac{K^\r}{2}\frac{\d}{i\bm e \,\d A_\l(x_3)} \frac{\d}{i\bm e\, \d A_\n(x_2)}\Big( A_\r \nabla_\m\braket{\cj^\m(x_1)} \Big)\\
&=\frac{K^\r}{2}\frac{\d}{i\bm e\, \d A_\l(x_3)} \frac{\d}{i\bm e \, \d A_\n(x_2)}\Big(A_\r \ca_{\rm chiral}(x_1) \Big).\numberthis \label{eqn:2-deriv}
\eala
The above procedures can be straightforwardly 
extended to the higher-point functions obtained by differentiating \eqref{eqn:3point-general} successively  with respect to the gauge fields, 
giving rise to 
constraints (putting together  \eqref{eqn:one-point-inconsistency}, \eqref{eqn:two-point-consistency} and \eqref{eqn:2-deriv} here) 
\begin{align}
0 & = K^\r A_\r \ca_{\rm chiral}(x), \label{eqn:0-deriv}\\
0 & = K^\r \frac{\d}{\bm e \, \d A_\n(x_1)}\Big( A_\r \ca_{\rm chiral}(x) \Big), \label{eqn:1-deriv} \\
0	& =  K^\r\frac{\d}{\bm e \, \d A_\l(x_2)} \frac{\d}{\bm e \, \d A_\n(x_1)}\Big( A_\r \ca_{\rm chiral}(x) \Big), \label{2-deriv} \\
0	&= K^\r \frac{\d}{\bm e \, \d A_\a(x_3)} \frac{\d}{\bm e\, \d A_\l(x_2)} \frac{\d}{\bm e \, \d A_\n(x_1)}\Big( A_\r \ca_{\rm chiral}(x) \Big), 
\label{eqn:higher-order-constraints} \\
0	&= K^\r \frac{\d}{\bm e \, \d A_\b(x_4)} \frac{\d}{\bm e \, \d A_\a(x_3)}  \frac{\d}{\bm e\, \d A_\l(x_2)} \frac{\d}{\bm e \, \d A_\n(x_1)}\Big(A_\r \ca_{\rm chiral}(x) \Big)
\label{eqn:4-deriv}
\end{align}
and so on. These  constraints  are consequences of the Ward identities  \eqref{eqn:cls-ward-susy}-\eqref{eqn:qt-ward-gauge} and relation \eqref{npt-X-Q} 
 with the supersymmetric vacuum condition \eqref{Q_z_h}. Constraint \eqref{eqn:4-deriv} and  the subsequent ones with the higher functional derivatives are trivially satisfied, 
 since $\ca_{\rm chiral}$ is quadratic in the gauge field, while 
 \eqref{eqn:0-deriv}-\eqref{eqn:higher-order-constraints} are nontrivial constraints.
In particular, 
constraint \eqref{eqn:higher-order-constraints} with \eqref{ca} gives
\begin{align}
0&= K^\r \frac{\d}{\bm e\, \d A_\a(x_3)}\frac{\d}{\bm e \, \d A_\l(x_2)}\frac{\d}{\bm e \, \d A_\n(x_1)}\Big( A_\r \ca_{\rm chiral}(x) \Big)\NO\\
&=2c_A\Big[K^\a\d^4(x,x_3)\ve^{\k\l\s\n}\pa_\k\d^4(x,x_2)\pa_\s\d^4(x,x_1)+\text{(permutations)}\Big],
\end{align}
which can be satisfied if and only if $c_A=0$. 
Thus,  the anomaly coefficient $c_A$ should vanish. This makes constraints  \eqref{eqn:0-deriv}-\eqref{2-deriv} satisfied automatically.

\section{Discussions 
}
\label{sec:discussion}
\setcounter{equation}{0}

In this note we have studied the quantum consistency 
of the new-minimal supergravity formulation of the $\cn=1$ supersymmetric theories with an $R$-symmetry in 3+1 dimensional curved space. By investigating the rigid supersymmetry algebra and the correlation functions obtained via differentiation   
of the Ward identities (with respect to the background gravitino and $R$-gauge fields),we have shown that 
the pure $U(1)_R$ chiral anomaly coefficient $c_A$ should vanish to be consistent with the supersymmetry. 
Our result indicates that the  supersymmetry is broken at the quantum level unless $c_A = 0$.

We emphasize that the anomaly coefficient 
$c_A$ depends only on the field content of the theory.
There exist some special cases where $c_A=0$. For instance, in the $\cn=1$ superconformal theories, the anomaly coefficient $c_A$ becomes
\begin{equation}
	c_A=5a-3c,
\end{equation}
where the central charges $a$ and $c$ (for free theory) is given by \cite{Christensen:1978gi, Anselmi:1997ys}
\begin{equation}
	a=\frac{1}{48}(9N_V+N_\c),\quad c=\frac{1}{24}(3N_V+N_\c).
\end{equation}
Here $N_V$ and $N_\c$ are the number of gauge and chiral multiplets, respectively. Therefore, $c_A$ becomes vanishing  
when $27 N_V=N_\c$, for example when $N_V=1$ and $N_\c=27$.
Another simple example for a theory with $c_A=0$ is the system that consists of a free chiral multiplet with $R$-charge 1 (see e.g. appendix \ref{app:free-chiral}). In this case, the Weyl fermion in the chiral multiplet is actually uncharged under the $U(1)_R$ symmetry and thus does not contribute to the $U(1)_R$ chiral anomaly.

In this work we have focused on the quantum consistency with respect to the pure $U(1)_R$ anomaly. We expect a similar result for the mixed $U(1)_R$ anomaly coefficient $c_m$. Namely, we anticipate  that higher-point correlation functions involving both of the supercurrent and the stress-energy tensor can be consistent only when $c_m=0$.  But
it needs more involved computations to show this, which 
we leave for the future work.

We have assumed that  
the supersymmetry Ward identity is non-anomalous, see \eqref{eqn:cls-ward-susy}, and our results bring forward a question about the supersymmetry anomaly, which is also related to the holography.  
In \cite{Papadimitriou:2017kzw,An:2017ihs} the holographic renormalization \cite{Henningson:1998gx,Balasubramanian:1999re,deBoer:1999tgo,Kraus:1999di,deHaro:2000vlm,Bianchi:2001de,Bianchi:2001kw,Martelli:2002sp,Skenderis:2002wp,Papadimitriou:2004ap} was carried out for both of the bosonic and fermionic sector of the 5D $\cn=2$  gauged supergravity, a virtual candidate for holographic dual of 4D $\cn=1$ superconformal field theory (that has an $R$-symmetry). 
By doing so, it was derived that the supersymmetric Ward identities for 4D $\cn=1$ superconformal field theories (SCFTs) contain anomaly-terms, which lead to the anomalous variation law of supercurrent operators under the rigid supersymmetry transformation.\footnote{Recently, a similar result was obtained in \cite{An:2018xxx}, where it was shown in the context of AdS$_3$/CFT$_2$ that in 2D $\cn=(1,1)$ superconformal field theories the supercurrent operator transforms anomalously under the rigid supersymmetry transformation.} 
If one assumes that the supersymmetry Ward identity \eqref{eqn:cls-ward-susy} receives anomaly corrections,  
those anomaly terms would introduce additional (contact) terms in the correlations functions considered in section \ref{SUSY} and \ref{sec:quantum-consistency}, which in turn may lead to complete cancellations on the right-hand sides of \eqref{eqn:0-deriv}-\eqref{eqn:4-deriv} restoring the consistency of the theory with $c_A\neq 0$.\footnote{One relevant question is whether there could exist local counterterms added to the action in such a way that the consistency of the theory is maintained without requiring $c_A=0$. In order for the local counterterms to change the argument of this paper and restore the consistency of the theory with $c_A\neq 0$, they should modify the Ward identities \eqref{eqn:cls-ward-susy}-\eqref{eqn:qt-ward-gauge} by introducing appropriate additional local terms to the Ward identities. However, the local counterterms that modify the Ward identities necessarily break the corresponding symmetries. Therefore, as far as one does not want to break supersymmetry and diffeomorphism invariance explicitly, we expect that local counterterms can not play a role in retrieving the consistency of the theory with $c_A\neq 0$.} 
It would be interesting to explore if the anomaly corrections obtained in \cite{An:2017ihs,Papadimitriou:2017kzw} by holographic renormalization could do the job (see \cite{Katsianis:2019hhg, Papadimitriou:2019gel} for recent field-theoretical studies on the supersymmetry anomalies). 
We hope to pursue this question in the future work.

\section*{Acknowledgments}

We thank  Gwang Il Kim and  Ui Ri Mun for useful discussions and comments. This research was supported in part by NSTC Project No. 130-01.

\appendix

\renewcommand{\thesection}{\Alph{section}}
\renewcommand{\theequation}{\Alph{section}.\arabic{equation}}

\section*{Appendices}
\setcounter{section}{0}

\section{Conventions}
\label{app:conventions}
\setcounter{equation}{0}

We follow the conventions used in \cite{Freedman:2012zz}. The metric signature is $(-,+,+,+)$ and $\ve_{0123}=1$. We denote $\g_5$ matrix by $\g_*$ in 4 dimensions, i.e.
\begin{equation}
	\g_*=i\g_0\g_1\g_2\g_3,
\end{equation}
which leads to useful formulas 
\bsal 
& \g^{\m\n\r}\g_*=i\ve^{\m\n\r\s}\g_\s,\\
& \ve^{\m\n\r\l}\g_{\r\l}=-2i\g^{\m\n}\g_*.
\esal
For any spinor $\c$, $\bar\c\equiv \c^T C$ is the Majorana conjugate of $\c$, where $C$ is the charge conjugation matrix. And all of the spinors in this note are Majorana ones.

In our conventions,
\bal
& \bar\l\G^{(r_1)}\cdots\G^{(r_p)}\c=t_0^{p-1}t_{r_1}\cdots t_{r_p}\bar\c\G^{(r_p)}\cdots\G^{(r_1)}\l,\\ & \c=\G^{(r_1)}\cdots\G^{(r_p)}\l\implies\bar\c=t_0^pt_{r_1}\cdots t_{r_p}\bar\l\G^{(r_p)}\cdots\G^{(r_1)},
\eal
where
\be 
t_0=t_3=1,\quad t_1=t_2=-1.
\ee 

Two kinds of connections appear in this note, i.e. the metric connection and the gauge connection. The symbol $\nabla$ stands for the covariant derivative with respect to the metric. The symbol $\cd$ indicates the connection with respect to the metric and the gauge field. For instance,
\be 
\cd_\m\j_\n\equiv \nabla_\m\j_\n+i G_\m\g_*\j_\n,\quad \cd_\m\bar\j_\n=\nabla_\m\bar\j_\n+i G_\m\bar\j_\n\g_*,
\ee 
where $G_\m\equiv A_\m-\frac32 V_\m$. Note that the supercurrent $\cs^\m$ is conjugate to the source $\j_\m$, and therefore it has a $U(1)_R$ charge opposite to $\j_\m$. It follows that one has to define the covariant derivative of $\cs^\m$ by
\begin{equation}
	\cd_\m\cs^\n=\nabla_\m\cs^\n-i G_\m\g_*\cs^\n.
\end{equation}

\section{An example: $\cn=1$ with a free chiral multiplet on $\mathbb{R}\times S_{r_3} $}\label{app:free-chiral}
\setcounter{equation}{0}

In this appendix we explicitly derive the variation of the current operators $\cs^\m$, $\cj^\m$ and $\cf^{\m\n}$ under the rigid supersymmetry transformation, for the special case when $\cn=1$ field theory for a free chiral multiplet is defined on $\mathbb R\times S_{r_3}$.

The action of the theory that we are interested in is given by \cite{Festuccia:2011ws},
\be \label{eqn:free-chiral-action}
S=\int \dd^4x\;\cl,
\ee 
where
\be \label{eqn:free-chiral-lagrangian}
\frac{1}{\bm e}\cl=FF^*-D_\m\f D^\m\f^*+i V^\m(\f^*D_\m\f-\f D_\m\f^*)-\frac q4(R+6V_\m V^\m)\f^*\f-\bar\j\g^\m \(D_\m-\frac{i}{2}V_\m\)P_L\j ,
\ee 
and $q$ is the $U(1)_R$-charge of the chiral multiplet, and
\bal 
& R=\frac{6}{r^2},\quad V_i=0,\;V_t=\frac{1}{r},\quad \; A_t=\frac{1}{r},\;A_i=0,\quad R_{\m\n}=2(V_\m V_\n-g_{\m\n}V_\r V^\r),\\
& D_\m\f\equiv \pa_\m\f-i qA_\m\f,
\quad D_\m\f^*\equiv \pa_\m\f+i qA_\m\f,
\quad D_\m\j\equiv \nabla_\m\j+i (q-1)A_\m\g_* \j,\\
&\nabla_\m\j\equiv \pa_\m\j+\frac14\o_{\m ab}\g^{ab}\j,\quad P_{L,R}\equiv\frac{1\mp\g_*}{2}.
\eal

This theory is invariant under the rigid supersymmetry
\bsub\label{eqn:free-chiral-susy-transformation}
\bal 
& \d\f=\bar\z P_L\j,\quad \d\f^*=\bar\z P_R\j,\\
& \d F= \bar\z\g^\m(D_\m-\frac{i}{2}V_\m)P_L\j,\quad \d F^*=\bar\z\g^\m\(D_\m+\frac{i}{2}V_\m\)P_R\j,\\
&\d P_L\j=P_L(\slashed D\f+F)\e,\quad \d P_R\j=P_R(\slashed D\f^*+F^*)\e,
\eal 
\esub
where $\z$ is the GKS that satisfies the GKS condition \eqref{eqn:GKS}. Note that under the charge conjugation $\g_*$ flips the sign and thus $(P_L)^C=P_R$.

The energy-momentum tensor is given by
\bala
\ct^{\m\n}=\;&-\frac{1}{2\bm e}\(e^{a\n}\frac{\d}{\d e^a_\m}+e^{a\m}\frac{\d}{\d e^a_\n}\)\cs \\
=\;&-g^{\m\n}\big[-D_\r\f D^\r\f^*+FF^*-\frac q4(R-6V^\r V_\r)\f^*\f  \big]-2D^{(\m}\f D^{\n)}\f^*+ \\
&+\frac q2(-R^{\m\n}+\nabla^\m \nabla^\n-g^{\m\n}\Box+6V^\m V^\n)(\f^*\f)-\frac{i}{2}V^{(\m}\bar\j\g^{\n)}\j\\
&+\frac{1}{2}[g^{\m\n}\bar\j\g^\r\overleftrightarrow D_\r\j-\bar\j\g^{(\m}\overleftrightarrow D^{\n)}\j],\numberthis
\eala 
and the $U(1)_R$-current is 
\be
\cj^\m=\frac{1}{\bm e}\frac{\d S}{\d A_\m} =i q(\f D^\m \f^*-\f^* D^\m\f)+2qV^\m\f\f^*+i(q-1)\bar\j\g^\m P_L\j .
\ee 
Since the Lagrangian \eqref{eqn:free-chiral-lagrangian} does not possess any FI-terms and the K\"{a}hler form of the target space is exact, there exists a well-defined operator $Y_\m$, such that $\cf_{\m\n}=\pa_\m Y_\n-\pa_\n Y_\m$, see e.g. \cite{Closset:2013vra}. Defining an operator $\ck^\m$ by
\be 
\ck^\m\equiv\frac{1}{\bm e}\frac{\d S}{\d V_\m}=i (\f^* D^\m\f-\f D^\m\f^*)-3q V^\m\f^*\f+\frac{i}{2}\bar\j\g^\m P_L\j,
\ee
we have an operator relation
\be\label{eqn:free-chiral-operator-relation} 
Y^\m= K^\m+\frac32\cj^\m=\frac{i}{2}(3q-2)(\f D^\m\f^*-\f^* D^\m\f+\bar\j\g^\m P_L\j)
\ee 
because
\be 
\d V_\m=\frac14\ve_\m{}^{\n\r\l}\pa_\n\d B_{\r\l}+(\d g_{\m\n})V^\n-\frac12 g^{\r\l}\d g_{\r\l} V_\m.
\ee
Notice that when $q=2/3$ the operator $Y^\m$ and $\cf_{\m\n}$ are identically vanishing. Since the conformal symmetry is explicitly broken by the operator $\cf_{\m\n}$ \cite{Klare:2012gn}, this implies that the theory becomes superconformal.

We could find the supercurrent by obtaining the Noether current corresponding to the transformation given by \eqref{eqn:free-chiral-susy-transformation}. Instead, we would like to use \eqref{eqn:susy-var-j-main} to find the supercurrent, which would differ from the Noether current by a term like $D_\n\cm^{\m\n}$. The variation of the $R$-current $\cj^\m$ is
\begin{align*}
	\d_\z\cj^\m&=-i\bar\z(\slashed D\f)\g^\m P_R\j+i q D_\n(\f\bar\j)\g^{\m\n}P_R\z+\frac{3q}{2}\f V^\m\bar\j P_R\z\\
	&\hskip5em+i q\f\bar\z\g^\m\g^\n[D_\n(P_R\j)+\frac{i}{2}V_\n P_R\j] +i(q-1)\bar\j\g^\m P_L\z F+\text{h.c.},
\end{align*}
so we find that
\begin{equation}
	\cs^\m=-i\bar\z(\slashed D\f)\g^\m P_R\j+i q D_\n(\f\bar\j)\g^{\m\n}P_R\z+\frac{3q}{2}\f V^\m\bar\j P_R\z+\text{h.c.},\label{eqn:supercurrent}
\end{equation}
where we used the equations of motion of the theory. Notice that although the rigid supersymmetry algebra is off-shell, the transformation rules \eqref{eqn:susy-var-j-main}, \eqref{eqn:susy-var-cf-main} and \eqref{eqn:supercurrent-transformation} should be on-shell relations at the classical level, as we see below.

The Gamma trace of the supercurrent \eqref{eqn:supercurrent} is
\begin{equation}
	\g_\m\cs^\m=-(3q-2)(\slashed D\f)P_R\j-3q\f\g^\n[D_\n(P_R\j)+\frac{i}{2}V_\n P_R\j]+\text{h.c.}.\label{eqn:gammatrace}
\end{equation}
This vanishes on-shell for $q=2/3$, which is related to the fact that 
the theory has the superconformal symmetry when $q=2/3$. It is also implied by \eqref{eqn:gammatrace} that \eqref{eqn:susy-var-cf-main} cannot hold off-shell, since $Y_\m$ vanishes identically for $q=2/3$.

One can see that \eqref{eqn:susy-var-cf-main} holds on-shell by observing that 
\begin{align*}
	\d_\z Y^\m&=\frac{i}{2}(3q-2)[\f D^\m(\bar\z P_R\j)-\bar\z P_R\j D^\m\f+\bar\j\g^\m P_L(\slashed D\f+F)\z]+\text{h.c.}\\
	&=\frac{i}{2}(3q-2)[D^\m(\f\bar\z P_R\j)-\bar\z\slashed D\f\g^\m P_R\j+\bar\j\g^\m P_L\z]+\text{h.c.} .
\end{align*}
By a tedious computation it can be explicitly shown that the transformation law for \eqref{eqn:supercurrent-transformation} also holds on-shell.  

We emphasize that the whole analysis in this appendix can be extended to the more general backgrounds that admit two supercharges with opposite $R$-charge.

\addcontentsline{toc}{section}{References}

\nocite{*}

\bibliographystyle{jhepcap}
\bibliography{susy_anomaly}

\end{document}